\documentclass[twocolumn,raggedbottom,longbibliography,floatfix]{revtex4-2}
\usepackage{amsmath,amssymb,bm}
\usepackage{graphicx}
\usepackage{hyperref}
\hypersetup{colorlinks=false,pdfborder={0 0 0}}

\begin{document}
\title{Helical Dirac Current with Local Coupling to a Chiral Potential}
\author{Ju Gao}
\email{jugao@illinois.edu}
\author{Fang Shen}
\affiliation{Department of Electrical and Computer Engineering, University of Illinois at Urbana--Champaign, Urbana, Illinois 61801, USA}
\date{October 8, 2026}

\begin{abstract}
We show that a propagating electron wave in the lowest cylindrical
channel carries a spin-resolved helical conserved current, and
connect this current texture to its local chiral coupling.
In a weakly relativistic positive-energy Dirac projection,
the two sectors have the same azimuthally symmetric charge density
and opposite circulation superposed on longitudinal flow,
even though their nonrelativistic orbital harmonic is $l=0$.
The current is radially distributed and extends into the
evanescent tail. A scalar screw potential couples these modes
through a derived vertex expressed by the \emph{chiral current
overlap}, $\mathcal J_\chi$. Both spatial textures determine
the coupling: the environmental radial profile weights the
electronic circulation, while the screw harmonic selects
correlated changes of spin sector and longitudinal momentum.
For a uniform static harmonic, elastic energy conservation
makes the first-order spin-flip amplitude vanish within the
same transverse mode. Environmental energy exchange is therefore
required to activate these selected channels. The same current
overlap supplies the spatial kernel for heliciton-assisted
inelastic scattering in a companion study and for a subsequent
transport treatment of chirality-induced spin selectivity (CISS).
\end{abstract}
\maketitle

\section{Introduction}
An electron wave can carry a spatially structured current even
when its charge density has no azimuthal winding. Orbital vortex
modes and externally shaped fields provide familiar routes to
helical flow~\cite{Verbeeck2010,Lloyd2017,Fang2024}. Spin also
contributes to the conserved Dirac current: its nonrelativistic
limit contains the established Pauli magnetization
current~\cite{Shikakhwa2011}. Here we resolve the helical current
of the lowest cylindrical channel and connect it explicitly
to local coupling with a scalar chiral potential. By
\emph{current texture} we mean the spatial organization of this
vector field.

In the wave--entity account~\cite{GaoShen2024_Entropy,GaoShen2025_JPhysComm},
the spatial Dirac wave carries charge and current as one
conserved physical four-current. For the modes considered here,
its longitudinal component describes propagation and its
azimuthal component expresses spin-associated circulation,
or wave--spin. The two sectors have the same azimuthally
symmetric charge density but opposite circulation. Their
current streamlines are helical, including in the evanescent
tail, although the nonrelativistic orbital harmonic is $l=0$.
The density alone therefore does not display the wave's full
spatial organization.

The local interaction depends on both the electronic current
and environmental potential textures. We derive the scalar
screw-potential vertex from the full Dirac modes and express
its radial factor through the \emph{chiral current overlap},
$\mathcal J_\chi$. The environmental radial profile samples
the existing circulation, while the screw harmonic selects
correlated spin-sector and longitudinal-momentum changes.
This gives the geometric bolt-and-nut scaffold a quantitative
meaning through a derived local overlap.

Chirality-induced spin selectivity (CISS) concerns spin-dependent
electron responses in chiral matter~\cite{NaamanWaldeck2015,YangEtAl2021}.
Physical scattering must also satisfy energy conservation.
For a uniform static screw
harmonic, elastic energy conservation makes the first-order
spin-flip amplitude vanish within the retained transverse mode.
Environmental energy exchange is therefore required to activate
these selected channels. A companion study represents this
exchange through a quantized environmental screw mode, whose
excitations are helicitons, and derives quantized environmental energy exchange and spin-resolved sidebands~\cite{GaoPaper2}.
A subsequent transport treatment must assemble the channel
populations and outgoing fluxes to predict net spin selectivity.
We follow the same spatial electron wave from its Dirac modes
to its conserved current, local chiral coupling, and the
energy-exchange conditions for conversion.

\section{Dirac modes in cylindrical confinement}
\label{sec:modes}
To resolve the electronic current texture and its local coupling,
we first construct the two propagating modes in cylindrical
confinement. The parent Dirac Hamiltonian is
\begin{equation}
\hat H_D=c\boldsymbol\alpha\cdot\mathbf p+\beta mc^2+U(\rho),
\quad \mathbf p=-i\hbar\nabla,\quad \beta=\gamma^0,
\label{eq:parent_hamiltonian}
\end{equation}
where $U(\rho)=0$ for $\rho<R$ and $U(\rho)=U>0$ for $\rho>R$.
The confinement is a potential energy multiplying the identity
in spinor space, as is the chiral perturbation introduced below.
We use positive-energy Dirac modes in a specified
weakly relativistic projection. For a stationary four-spinor
$\Psi=(\varphi,\nu)^T$, solving for the lower component gives
\begin{equation}
\nu=[\epsilon_k-U(\rho)+mc^2]^{-1}
c\,\boldsymbol\sigma\cdot\mathbf p\,\varphi
\simeq\frac{\boldsymbol\sigma\cdot\mathbf p}{2mc}\varphi.
\label{eq:pauli_reconstruction}
\end{equation}
The inverse factor acts locally after
$\boldsymbol\sigma\cdot\mathbf p$. We retain the leading
confined radial function and this reconstruction when evaluating
the full current and scalar matrix element. Relativistic changes
of the radial modes, including confinement-induced spin mixing,
are excluded. Thus the vertex is a result within this projection,
not a complete order-$c^{-2}$ calculation for exact finite-step
Dirac eigenstates.

Let $E_\perp$ be the \emph{transverse confinement energy} of the
lowest radial mode in this approximation. It is the eigenvalue
of the transverse confinement problem and excludes rest energy
and longitudinal kinetic energy:
\begin{equation}
\left[-\frac{\hbar^2}{2m}\frac{1}{\rho}\partial_\rho
\bigl(\rho\partial_\rho\bigr)+U(\rho)\right]F(\rho)
=E_\perp F(\rho),
\label{eq:transverse_eigenproblem}
\end{equation}
\begin{equation}
\epsilon_k=mc^2+E_\perp+\frac{\hbar^2k^2}{2m}.
\label{eq:E_k}
\end{equation}
For $0<E_\perp<U$, define the interior oscillation wave number
$\zeta$ and exterior decay constant $\xi$ by
\begin{equation}
\zeta^2=\frac{2mE_\perp}{\hbar^2},\quad
\xi^2=\frac{2m(U-E_\perp)}{\hbar^2}.
\label{eq:radial_parameters}
\end{equation}
The radial functions are
\begin{equation}
F(\rho)=\begin{cases}J_0(\zeta\rho),&\rho\le R,\\
\kappa K_0(\xi\rho),&\rho>R,\end{cases}
\quad G(\rho)=-\frac{dF}{d\rho},
\label{eq:radial_functions}
\end{equation}
so $G=\zeta J_1$ inside and $G=\kappa\xi K_1$ outside.
Continuity of $F$ and $F'$ yields
\begin{equation}
\kappa=\frac{J_0(\zeta R)}{K_0(\xi R)},\quad
\zeta\frac{J_1(\zeta R)}{J_0(\zeta R)}
=\xi\frac{K_1(\xi R)}{K_0(\xi R)}.
\label{eq:radial_matching}
\end{equation}
This matching condition defines the radial eigenvalue in the
stated approximation. The same reconstruction coefficient
\begin{equation}
\eta=\frac{\hbar}{2mc}
\label{eq:eta_k}
\end{equation}
is used in both regions. The working regime requires
$E_\perp,U,\hbar^2k^2/(2m)\ll mc^2$ and
$\eta\zeta,\eta\xi,\eta|k|\ll1$; selected final momenta must
also remain in this regime.

Write the modes as $\Psi_{sk}=u_{sk}(\rho,\phi)e^{ikz}$, with
\begin{equation}
u_{\uparrow k}=N_k
\begin{pmatrix}F\\0\\\eta kF\\i\eta G e^{i\phi}\end{pmatrix},
\label{eq:wavefunction_up}
\end{equation}
\begin{equation}
u_{\downarrow k}=N_k
\begin{pmatrix}0\\F\\i\eta G e^{-i\phi}\\-\eta kF\end{pmatrix}.
\label{eq:wavefunction_down}
\end{equation}
\newpage
Here $N_k$ is real and positive. We normalize the full
transverse four-spinor to unity:
\begin{equation}
\begin{aligned}
\int d^2\rho\,u_{s'k}^\dagger u_{sk}&=\delta_{s's},\\
N_k^{-2}&=2\pi\int_0^\infty
\left[(1+\eta^2k^2)F^2+\eta^2G^2\right]\rho\,d\rho.
\end{aligned}
\label{eq:transverse_normalization}
\end{equation}
Including the reconstructed terms in this normalization does
not include the omitted corrections to the radial modes.
The longitudinal convention is
\begin{equation}
\langle\Psi_{s'k'}|\Psi_{sk}\rangle
=2\pi\delta_{s's}\delta(k'-k),
\label{eq:continuum_normalization}
\end{equation}
with integration measure $dk/(2\pi)$. A normalized wavepacket
has the form
$\Psi=\sum_s\int dk\,a_s(k)\Psi_{sk}/(2\pi)$, where
$\sum_s\int dk\,|a_s(k)|^2/(2\pi)=1$.

The label $l=0$ specifies the channel's orbital harmonic in
its nonrelativistic limit. The angular factors $e^{\pm i\phi}$
in the full Dirac modes are compatible with their definite
total angular momentum:
\begin{equation}
\hat J_z\Psi_{\uparrow k}=\frac{\hbar}{2}\Psi_{\uparrow k},\quad
\hat J_z\Psi_{\downarrow k}=-\frac{\hbar}{2}\Psi_{\downarrow k},
\label{eq:Jz_sectors}
\end{equation}
where $\hat J_z=-i\hbar\partial_\phi+(\hbar/2)\Sigma_z$
and $\boldsymbol\Sigma=\operatorname{diag}
(\boldsymbol\sigma,\boldsymbol\sigma)$.
The full four-spinors are not zero-eigenvalue states of
$\hat L_z$. The arrows label spin sectors connected to
$\Sigma_z=\pm1$ in the nonrelativistic limit; the full modes
are not exact $\Sigma_z$ eigenstates. Below, spin-flip denotes
conversion between these two sectors.

\section{Helical geometry of the full Dirac current}
The physical charge current of an electron wave is the
Lorentz-covariant Dirac bilinear
\begin{equation}
j^\mu=-ec\,\bar\Psi\gamma^\mu\Psi
=(c\rho_e,\mathbf j),\quad e>0,
\label{eq:dirac_current}
\end{equation}
with $\bar\Psi=\Psi^\dagger\gamma^0$,
$\rho_e=-e\Psi^\dagger\Psi$, and
$\mathbf j=-e\Psi^\dagger(c\boldsymbol\alpha)\Psi$.
Charge density and spatial current are components of the same
four-current, conserved by Dirac dynamics, $\partial_\mu j^\mu=0$.
We evaluate it from the complete spinor and the Dirac velocity
operator. All geometry below is described in the rest frame
of the confinement.

For the continuum modes, denote the transverse mode current by
$j_a^s(\rho;k)=-ec\,u_{sk}^\dagger\alpha_{\hat a}u_{sk}$.
Its physical cylindrical components use
\begin{equation}
\alpha_{\hat\rho}=\alpha_x\cos\phi+\alpha_y\sin\phi,\quad
\alpha_{\hat\phi}=-\alpha_x\sin\phi+\alpha_y\cos\phi,
\label{eq:cylindrical_alpha}
\end{equation}
and $\alpha_{\hat z}=\alpha_z$. In particular, $j_\phi$
is the component along $\hat{\boldsymbol\phi}$.
The modal current is defined per unit longitudinal wave
amplitude; for a monochromatic channel with electron number
per unit length $n_z$, the physical current density is
$n_z\mathbf j^s$. Finite-wavepacket currents follow by applying
Eq.~\eqref{eq:dirac_current} to the normalized superposition
in Sec.~\ref{sec:modes}.
This convention removes any auxiliary box length.

Direct evaluation gives
\begin{equation}
j_\rho^\uparrow=j_\rho^\downarrow=0,
\label{eq:j_rho}
\end{equation}
\begin{equation}
j_z^\uparrow(\rho;k)=j_z^\downarrow(\rho;k)
=-2ecN_k^2\eta kF^2(\rho),
\label{eq:j_z}
\end{equation}
\begin{equation}
j_\phi^\uparrow(\rho;k)=-2ecN_k^2\eta F(\rho)G(\rho),\quad
j_\phi^\downarrow(\rho;k)=-j_\phi^\uparrow(\rho;k).
\label{eq:j_phi_up}
\end{equation}
Inside, $FG=\zeta J_0J_1$; outside,
$FG=\kappa^2\xi K_0K_1$. Since $c\eta=\hbar/(2m)$,
the leading currents remain finite in the nonrelativistic limit.
The mode density
$u_{sk}^\dagger u_{sk}=N_k^2[(1+\eta^2k^2)F^2+\eta^2G^2]$
is identical in the two sectors and azimuthally symmetric.
Their opposite circulation is thus a distinction in the physical
vector current that the common charge density does not display.
The same current combines traveling flow with spin-associated
circulation, giving helical streamlines for $k\ne0$ without
an applied helical field or an orbital vortex in the $l=0$ limit.
Each stationary modal current is divergence-free: its radial
component vanishes and its other components are independent
of $\phi$ and $z$.

A current streamline is an integral curve of this vector field,
satisfying $d\mathbf r\parallel\mathbf j$.
At fixed radius and for $k\ne0$,
\begin{equation}
\frac{d\phi}{dz}=\frac{j_\phi^s}{\rho j_z^s}
=s\frac{G(\rho)}{\rho kF(\rho)},\quad s=+1,-1,
\label{eq:helix_streamline}
\end{equation}
for the up and down sectors, respectively. Integration gives
the azimuthal coordinate $\phi_s(\rho;z)$ of a current streamline
at fixed radius $\rho$:
\begin{equation}
\phi_s(\rho;z)=\phi_0+s\frac{\zeta J_1(\zeta\rho)}
{\rho kJ_0(\zeta\rho)}z,\quad 0<\rho<R,
\label{eq:helix_solution_inside}
\end{equation}
\begin{equation}
\phi_s(\rho;z)=\phi_0+s\frac{\xi K_1(\xi\rho)}
{\rho kK_0(\xi\rho)}z,\quad \rho>R.
\label{eq:helix_solution_outside}
\end{equation}
Here $\phi_0$ is the azimuthal coordinate at $z=0$.

The streamlines describe the organization of the distributed
current; they do not specify individual point-electron histories.
The term helical current refers to this streamline geometry,
rather than an eigenvalue of the conventional spin--momentum
helicity operator. At fixed propagation direction, the spin
sectors have opposite handedness. Reversing $k$ also reverses
the handedness; at $k=0$, the current is purely azimuthal and
the streamlines are circles. Although the exterior amplitudes
decay, their ratio continues to define a local pitch in the
evanescent tail.

For $k>0$, the positive axial pitch in the interior is
\begin{equation}
Z(\rho)=\frac{2\pi}{|d\phi/dz|}
=2\pi\frac{\rho kJ_0(\zeta\rho)}{\zeta J_1(\zeta\rho)},
\quad 0<\rho<R,
\label{eq:Z_def}
\end{equation}
for the lowest radial mode, where $J_0$ and $J_1$ are positive.
This geometric pitch measures current orientation and is
distinct from the axial phase wavelength $\lambda_z=2\pi/k$.
At fixed confinement and radius, $Z(\rho)\propto k$ whereas
$\lambda_z\propto k^{-1}$; both are determined by the prepared mode.

\begin{figure}[t]
\centering
\includegraphics[width=0.72\linewidth]{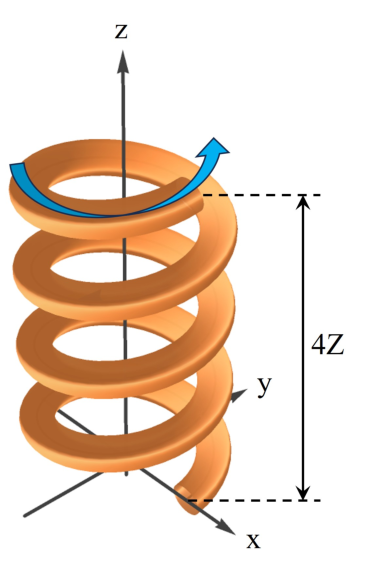}
\caption{Representative helical current streamline for the up
sector at $k>0$, passing through $\rho^*$ where the azimuthal
current magnitude is largest. The label $4Z$ spans four
geometric pitches. The tube provides visual thickness; the
current also occupies other radii. Reversing the spin sector
reverses the handedness at fixed $k$. The blue arrow indicates the direction of electron number flow, opposite to the electric-current direction because the electron charge is negative.}
\label{fig:helical_current}
\end{figure}

To illustrate this geometry, choose the radius of maximal
azimuthal current magnitude. An interior maximum satisfies
\begin{equation}
\frac{d|j_\phi^\uparrow|}{d\rho}=0,
\label{eq:jphi_max}
\end{equation}
which gives
\begin{equation}
J_0(\zeta\rho^*)[J_0(\zeta\rho^*)-J_2(\zeta\rho^*)]
=2J_1^2(\zeta\rho^*).
\label{eq:rho_star_condition}
\end{equation}
For $R=1\,\mathrm{nm}$, $U=2\,\mathrm{eV}$, and longitudinal
kinetic energy $E_\parallel=\hbar^2k^2/(2m)=25\,\mathrm{meV}$,
Eq.~\eqref{eq:radial_matching} yields
$E_\perp=169.50\,\mathrm{meV}$ and
$\zeta=2.1092\,\mathrm{nm}^{-1}$. With
$k=0.81004\,\mathrm{nm}^{-1}$, the representative values are
\begin{equation}
\begin{aligned}
\rho^*&=0.51297\,\mathrm{nm},\\
Z(\rho^*)&=1.9356\,\mathrm{nm},\quad
\lambda_z=7.7566\,\mathrm{nm}.
\end{aligned}
\label{eq:Z_rep}
\end{equation}
Figure~\ref{fig:helical_current} illustrates the up-sector
current streamline through $\rho^*$ with $\phi_0=0$.
The electron wave and its current
remain distributed over the radial profile, including the
evanescent tail.

\section{Local chiral coupling and the current overlap}

Having resolved the electronic current texture, we specify
the environmental texture through a scalar potential energy
with a single screw harmonic,
\begin{equation}
V_\chi(\rho,\phi,z)=
\begin{cases}
V_0f(\rho)g(z)\cos(\phi-qz),&0\le\rho<R,\\
0,&\rho\ge R.
\end{cases}
\label{eq:Vchi}
\end{equation}
Here $V_0$ has units of energy, $f$ and $g$ are dimensionless
real profiles, and $q=2\pi/p$ specifies the signed screw
wave number and pitch magnitude $|p|$. We use $g(z)=1$ for
the uniform harmonic and address a finite envelope below.
The potential is an idealization of a helical electrostatic
modulation produced, for example, by a surrounding molecular
structure or patterned gates. The restriction to interior
support is a model choice.

A regular first angular harmonic requires $f(\rho)=O(\rho)$
at the axis. One admissible near-wall profile is
\begin{equation}
f(\rho)=\frac{\rho}{R}
\exp\!\left[-\frac{(\rho^2-R^2)^2}{8R^2a^2}\right],
\quad 0<a<R,
\label{eq:radial-profile}
\end{equation}
whose exponential approaches a Gaussian of width $a$ near
$\rho=R$. The results below hold for any such radial profile.

We evaluate the perturbation to first order in $V_0$ within
the retained modes. The scalar interaction $V_\chi$ defines
the continuum coupling vertex matrix $\boldsymbol M(k',k)$.
Its rows label the final spin sector $s'$ and its columns
the initial spin sector $s$, in the order
$(\uparrow,\downarrow)$, with elements
\begin{equation}
M_{s's}(k',k)
=\langle\Psi_{s'k'}|V_\chi|\Psi_{sk}\rangle.
\label{eq:matrix-def}
\end{equation}
The scalar interaction couples the wave and potential at each
spatial point; projection integrates this local coupling over
the extended modes. These matrix elements are evaluated before
imposing an energy-conservation condition.

For the initial up sector and final down sector, direct
evaluation with the full spinors gives, for $g(z)=1$,
\begin{equation}
\begin{aligned}
M_{\downarrow\uparrow}(k',k)
={}&\frac{i\eta}{2ec}\frac{N_{k'}}{N_k}(k+k')\\
&\times\int d^3r\,V_\chi(\mathbf r)\,
j_\phi^\uparrow(\rho;k)e^{i\phi+i(k-k')z}\\
={}&-i\frac{\pi V_0\eta}{2ec}
\frac{N_{k'}}{N_k}(k+k')\mathcal J_\chi(k)\\
&\times 2\pi\delta(k'-k-q),
\end{aligned}
\label{eq:opposite_spin_overlap}
\end{equation}
where the chiral current overlap is
\begin{equation}
\begin{aligned}
\mathcal J_\chi(k)
&=-\int_0^R f(\rho)\,
j_\phi^\uparrow(\rho;k)\,\rho\,d\rho\\
&=2ecN_k^2\eta\zeta
\int_0^R f(\rho)J_0(\zeta\rho)J_1(\zeta\rho)\,
\rho\,d\rho.
\end{aligned}
\label{eq:chiral_current_overlap}
\end{equation}

Equation~\eqref{eq:opposite_spin_overlap} makes both spatial
textures explicit in the scalar matrix element. The environmental
profile $f(\rho)$ weights the electron's circulating current
through $\mathcal J_\chi$, while the screw harmonic supplies
angular and longitudinal selection. Both textures therefore
determine the strength and selection of the conversion links;
the common charge density alone does not specify this kernel.
Under the transverse normalization used here,
$\mathcal J_\chi$ has units of current times length.
If the same harmonic also has exterior support, its contribution
is obtained by extending the overlap integral into the
evanescent current.

\subsection{Selected vertices and momentum transfer}

For $g(z)=1$, the potential has two screw components,
\begin{equation}
V_\chi^{(\pm)}
=\frac{V_0f(\rho)}{2}e^{\pm i\phi\mp iqz},
\quad 0\le\rho<R.
\label{eq:Vchi_plus}
\end{equation}
In same-sector overlaps, the angular phases cancel, leaving
only the zero angular harmonic. Neither screw component
then survives angular integration, so
\begin{equation}
M_{\uparrow\uparrow}(k',k)
=M_{\downarrow\downarrow}(k',k)=0.
\label{eq:diagonal-zero}
\end{equation}
This angular selection holds for the first-order projection
onto the retained modes, not for every spin-preserving
process in a multimode scattering problem.

For opposite sectors, angular integration selects
$V_\chi^{(-)}$ for $\uparrow\to\downarrow$ and
$V_\chi^{(+)}$ for the reverse link. The longitudinal
integral gives $2\pi\delta(k'-k\mp q)$.
Define the reduced vertices $\mathcal V_{s's}(k',k)$
by separating these momentum selectors from the full
continuum matrix elements:
\begin{equation}
\begin{aligned}
M_{\downarrow\uparrow}(k',k)
&=2\pi\delta(k'-k-q)\,
\mathcal V_{\downarrow\uparrow}(k',k),\\
M_{\uparrow\downarrow}(k',k)
&=2\pi\delta(k'-k+q)\,
\mathcal V_{\uparrow\downarrow}(k',k),
\end{aligned}
\label{eq:matrix2}
\end{equation}
where
\begin{equation}
\begin{aligned}
\mathcal V_{\downarrow\uparrow}(k',k)
&=-i\frac{\pi V_0\eta}{2ec}
\frac{N_{k'}}{N_k}(k+k')\mathcal J_\chi(k),\\
\mathcal V_{\uparrow\downarrow}(k',k)
&=+i\frac{\pi V_0\eta}{2ec}
\frac{N_{k'}}{N_k}(k+k')\mathcal J_\chi(k).
\end{aligned}
\label{eq:M-down-up-current}
\end{equation}
The full matrix elements satisfy
$M_{s's}(k',k)=M_{ss'}^*(k,k')$.
The reduced vertices have units of energy, whereas the
continuum matrix elements include the dimensionful
momentum delta. Both the selector and the measure
$dk/(2\pi)$ must be retained when constructing scattering
amplitudes or rates.

The momentum selectors give the conversion rules
\begin{equation}
\uparrow,k\longrightarrow\downarrow,k+q,\quad
\downarrow,k\longrightarrow\uparrow,k-q.
\label{eq:momentum-selection}
\end{equation}
In terms of $m_J=\pm1/2$, the eigenvalue of $\hat J_z/\hbar$, they read
$\Delta m_J=\pm1$ and $\Delta k=-q\Delta m_J$.
Reversing the screw handedness reverses this correlation
between spin-sector change and momentum transfer.

On these two selected links, the reduced vertices can
be collected in the array
\begin{equation}
\begin{aligned}
\boldsymbol{\mathcal V}_{\rm sel}(k)
={}&i\frac{\pi V_0\eta}{2ec}\mathcal J_\chi(k)\\
&\times\begin{pmatrix}
0&\dfrac{N_{k-q}}{N_k}(2k-q)\\[5pt]
-\dfrac{N_{k+q}}{N_k}(2k+q)&0
\end{pmatrix}.
\end{aligned}
\label{eq:matrix3}
\end{equation}
The array collects two conversion links with different
final momenta. At the leading order of this spin-flip
vertex, $N_{k\pm q}/N_k$ may be set to unity; the
normalization changes give higher-order corrections
within the retained basis.

\subsection{Elastic constraint and dynamical extension}
\label{sec:dynamics}

The selected vertex supplies the first Born kernel. We impose
energy conservation to identify the dynamical extension required
for nonzero scattering in the uniform, same-mode model.

For a static environment, electron-elastic scattering within
the same transverse mode requires
\begin{equation}
\epsilon_{k'}=\epsilon_k
\quad\Longleftrightarrow\quad k'^2=k^2.
\label{eq:elastic_energy_condition}
\end{equation}
With $g(z)=1$ and $q\ne0$, the momentum selector excludes
$k'=k$. The only possible elastic link is therefore
$k'=-k$, where the vertex factor $k+k'$ vanishes:
\begin{equation}
\mathcal V_{\rm flip}^{\rm elastic}=0.
\label{eq:elastic_zero}
\end{equation}
Thus the unequal off-shell factors $2k\pm q$ do not
establish elastic spin filtering.

For a finite envelope, $2\pi\delta(k'-k\mp q)$ becomes
$\widetilde g(k'-k\mp q)$, with
$\widetilde g(Q)=\int dz\,g(z)e^{-iQz}$.
Exact first-order backscattering still vanishes; elastic
forward mixing can occur but does not alone establish
spin filtering.

A nonzero selected uniform conversion must therefore be
electron-inelastic: the environment participates in the
energy and momentum balance. Recoil or internal excitation
can provide this dynamical response; a prescribed
time-dependent modulation can also supply energy.
A companion study~\cite{GaoPaper2} describes this exchange
through a quantized environmental screw mode, whose
excitations are helicitons with energy $\hbar\Omega_q>0$.
For one reference handedness, the uniform absorption and
emission channels are
\begin{equation}
\begin{aligned}
|\uparrow,k;n_q\rangle
&\longrightarrow|\downarrow,k+q;n_q-1\rangle,\\
|\downarrow,k;n_q\rangle
&\longrightarrow|\uparrow,k-q;n_q+1\rangle,
\end{aligned}
\label{eq:heliciton_channels}
\end{equation}
where $n_q$ is the mode occupation number.
Total electron--mode energy is conserved under the
respective resonance conditions
\begin{equation}
\begin{aligned}
\epsilon_{k+q}-\epsilon_k
&=\frac{\hbar^2q(2k+q)}{2m}=\hbar\Omega_q,\\
\epsilon_k-\epsilon_{k-q}
&=\frac{\hbar^2q(2k-q)}{2m}=\hbar\Omega_q.
\end{aligned}
\label{eq:inelastic_energy_conditions}
\end{equation}
These conditions make the corresponding vertex factors
$2k\pm q$ nonzero. Quantization supplies the absorption
and emission factors $\sqrt{n_q}$ and $\sqrt{n_q+1}$,
while $\mathcal J_\chi$ retains the spatial coupling.

Absorption and emission change the environmental occupation
as well as the electronic sector, making the environment a
dynamical participant in the conversion. The changed occupation
can persist until further exchange or relaxation, so subsequent
conversion probabilities can depend on the state left by earlier
interactions.

Using the same current-overlap structure, the companion study
develops quantized environmental energy exchange and spin-resolved
sidebands. A subsequent scattering and transport treatment must
establish flux-normalized conversion probabilities and assemble
the outgoing spin-resolved flux.

\section{Discussion}
\label{sec:discussion}

We present a wave--entity account of a helical electron wave
in the lowest cylindrical channel, whose nonrelativistic orbital
harmonic is $l=0$, locally coupled to a chiral environment.
The identical charge densities of its two sectors coexist
with opposite circulating currents. This radially distributed
structure becomes quantitatively relevant to their interaction:
Eq.~\eqref{eq:opposite_spin_overlap} expresses the scalar
conversion matrix element through the azimuthal current,
and $\mathcal J_\chi$ specifies its environmental sampling.

The projected vertex is derived from the Dirac modes and
specified scalar potential. At the retained order, it also
follows from the spin--orbit coupling (SOC) term in the
Foldy--Wouthuysen reduction~\cite{FoldyWouthuysen1950},
projected onto the same modes with the potential's boundary
terms retained. The current remains finite in the
nonrelativistic limit, while the scalar conversion vertex
is of order $c^{-2}$. The current-overlap form makes the
electronic circulation and its local sampling explicit for
a continuous screw potential, complementing
curved-path~\cite{Shitade2020} and helical
atomic-scatterer~\cite{VanRuitenbeek2023} treatments.
The kernel belongs to the stated mode projection;
confinement-mode corrections can contribute at the same
relativistic order.

A related spatial account appears in the angular-momentum-resolved
Aharonov--Bohm response~\cite{GaoAB2026}, where the first-order
vector-potential coupling energy is
$-\int\mathbf j\cdot\mathbf A\,d^3r$.
Here a scalar screw potential produces sector conversion
whose matrix element exposes the circulating-current texture.
The interaction expressions and responses differ, but both
show how an external potential samples the distributed
Dirac-wave structure.

Spatial matching establishes a coupling kernel; energy
conservation determines which transitions it can produce.
The uniform same-mode elastic zero requires environmental
energy exchange to activate the selected first-order links.
Recoil or internal excitation can provide a dynamical partner,
with total energy conserved by the combined system.
Heliciton absorption and emission give a concrete quantum
description of this exchange while retaining the same
spatial current overlap.

The resulting CISS program asks how a spatially structured
electron wave exchanges energy and momentum locally with a
chiral environment, and how the conversion changes its
outgoing spin-resolved flux. The present work supplies the
current geometry and local kernel. A companion study develops
heliciton-assisted inelastic scattering~\cite{GaoPaper2},
while a subsequent transport treatment will assemble the
channel populations and fluxes into a prediction of spin
selectivity. That complete scattering description must include
all transmitted and reflected channels under specified
incident and environmental populations. Material applications
also require the environmental profile and spectrum.

\section{Conclusion}
\label{sec:conclusion}

In the wave--entity account, the spatial electron wave carries
charge, traveling flow, and spin-associated circulation as
one physical conserved-current structure. The derived chiral
current overlap gives its spin--chiral interaction an explicit
local meaning: the environmental potential samples the
electron's distributed circulation, and both spatial textures
determine the conversion kernel. The extended electron wave
thus participates in a locally mediated spin--chiral interaction
through its physical current texture.

\begin{acknowledgments}
The authors thank Prof.\ Wei Wang for stimulating discussions
on spin and chirality in biological and biophysical systems.
\end{acknowledgments}
% References are in Helix.bib; run BibTeX between the first and second XeLaTeX passes.
% REVTeX selects apsrev4-2 and supplies the longbibliography controls.
\bibliography{Helix}

\end{document}